\documentclass{article}
\usepackage{ijcai20}

\usepackage{times}
\usepackage{soul}
\usepackage{url}
\usepackage[hidelinks]{hyperref}
\usepackage[utf8]{inputenc}
\usepackage[small]{caption}
\usepackage{graphicx}
\usepackage{amsmath}
\usepackage{amsthm}
\usepackage{booktabs}
\usepackage{amsfonts}
\usepackage[ruled,linesnumbered]{algorithm2e}
\usepackage{subfigure}

\newcommand{\la}{\leftarrow}

\title{FPGA-Based Hardware Accelerator of Homomorphic Encryption for Efficient Federated Learning}

\author{
Zhaoxiong Yang$^1$\and
Shuihai Hu$^2$\and
Kai Chen$^{1,3}$
\affiliations
$^1$SING Lab, Hong Kong University of Science and Technology\\
$^2$Clustar, $^3$Peng Cheng Lab\\
\emails
zyangas@connect.ust.hk,
shuihai@clustar.ai,
kaichen@cse.ust.hk
}
\begin{document}

\maketitle

\begin{abstract}
  With the increasing awareness of privacy protection and data fragmentation problem, federated learning has been emerging as a new paradigm of machine learning. Federated learning tends to utilize various privacy preserving mechanisms to protect the transferred intermediate data, among which homomorphic encryption strikes a balance between security and ease of utilization. However, the complicated operations and large operands impose significant overhead on federated learning. Maintaining accuracy and security more efficiently has been a key problem of federated learning. In this work, we investigate a hardware solution, and design an FPGA-based homomorphic encryption framework, aiming to accelerate the training phase in federated learning. The root complexity lies in searching for a compact architecture for the core operation of homomorphic encryption, to suit the requirement of federated learning about high encryption throughput and flexibility of configuration. Our framework implements the representative Paillier homomorphic cryptosystem with high level synthesis for flexibility and portability, with careful optimization on the modular multiplication operation in terms of processing clock cycle, resource usage and clock frequency. Our accelerator achieves a near-optimal execution clock cycle, with a better DSP-efficiency than existing designs, and reduces the encryption time by up to 71\% during training process of various federated learning models.
  
\end{abstract}

\section{Introduction}

In recent years, deep learning has made an unprecedented leap in the ability of human discovering knowledge and comprehending the world. Nevertheless, the adoption of deep learning is now faced with two barriers, namely data fragmentation and privacy preservation\cite{yang2019federated}. Federated learning has come up as a new machine learning paradigm to tackle the issues, learning models from decentralized datasets in a secure way.

To preserve data privacy, federated learning usually employs various mechanisms like differential privacy (DP), homomorphic encryption (HE), secure multiparty computation (SMC), etc. Whereas DP does not prevent data leakage completely, and the intricate protocols that SMC introduces to the system renders it virtually impractical, HE achieves a balance between security and operability. Moreover, one of the HE scheme named Paillier encryption scheme\cite{paillier1999public-key} has been adopted to protect the data privacy in neural networks\cite{ma2017secure}, logistic regression\cite{hardy2017private}, Bayesian network\cite{wright2004privacy-preserving}, clustering\cite{bunn2007secure}, showcasing its great generality as a privacy preserving mechanism in machine learning. 

However, the complicated operations and large operands of HE still impose overhead on federated learning that cannot be ignored. Research community and industry have been haunted by the question of how to provide \emph{secure}, \emph{accurate}, and yet \emph{efficient} federated learning. Previous effort such as FATE\cite{FATE}, a cutting edge federated learning system, has provided convenient interface to implement learning algorithms secured by Paillier HE, but the learning throughput is limited due to encryption by software.
In this work, we seek for a hardware solution to improve the training throughput of federated learning, designing a homomorphic encryption framework based on FPGA, since FPGA acceleration card has been commonly available in datacenters\cite{putnam2014large} and usually achieve a lower power consumption than GPU. The framework devises a customized FPGA implementation of the Paillier homomorphic encryption, and provides support for federated learning models with secure information exchange. 
%As far as we know, there has not been research works on hardware acceleration for cryptographic computation in the context of federated learning.

% TODO: rewrite the challenge part
%The main challenge lies in how to make full use of the flexibility and parallel computing resource of FPGA for acceleration. On the one hand, computation and storage resource on an FPGA is limited. To deliver a maximum throughput of encryption, it is desirable to make each encryption core resource-efficient, balancing execution time and resource consumption. On the other hand, computational units on FPGA are diversified with each configurable, while the overall throughput could be affected by intertwining metrics, e.g., a component reducing the execution clock cycle may also lower the clock frequency. Hence looking for an optimal configuration for the best throughput could be demanding.
%As the Paillier cryptosystem is composed of intensive arithmetic operations, and digital signal processors are the high performance units on FPGA performing arithmetic, we concentrate on maximizing the DSP-efficiency, namely throughput provided by each DSP. Nevertheless, we are not ignorant of the area and memory usage, and ensure not leaving them bottlenecks. 
%Last but not least, a peripheral accelerator inevitably introduces overhead of data transfer or kernel invocation. Eliminating or mitigating the overhead will be necessary for a good acceleration ratio compared with host CPU.

We demonstrate in this work that homomorphic encryption is usually composed of iterative operations that are hard to parallelize. Therefore, it is more reasonable to consider parallelism across data items to be encrypted, and make each encryption core compact and resource efficient, so as to maximize the overall throughput to handle the massive data in federated learning. The existing works fail to do that, as they either try to exhaust the resource on a single FPGA chip to produce one encryption unit to minimize the processing latency, or they mainly utilize the common circuit units (usually termed CLB or LUT) without making use of the digital signal processing (DSP) units, which are the powerful units for high performance arithmetic operation on modern FPGA. Moreover, most of them rely on the traditional register-level transfer (RTL) approach, lacking the flexibility of fast development and reconfiguration. 
In this work, we base our design and implementation on high level synthesis (HLS) that describe the FPGA circuit with high-level programming language for flexibility, allowing the algorithm and operations to be parametric and portable, and we try to derive an analytical model that determines the encryption performance, carry out optimization from multiple dimension.

%In this work, we base our design and implementation on high level synthesis (HLS) for flexibility, allowing the algorithm and operations to be parametric and portable. 
Since the bulk of computation of Paillier cryptosystem boils down to modular multiplication(ModMult), we focus on designing compact architecture for ModMult operation. We adopt the Montgomery algorithm\cite{montgomery1985modular} to carry out the operation, which is FPGA-friendly as it eliminates integer division operations. We figure out the key factors that determines the total en/decryption throughput on an FPGA chip, conduct overall optimization on Paillier processors in terms of clock cycle, resource consumption, clock frequency and memory usage respectively to attain the best throughput.

%To address resource limitation, note that the Paillier cryptosystem is composed of intensive arithmetic operations, we mainly employ DSP units to perform the computation and focus on its utilization efficiency. The other circuit units are responsible for control logic, and we take care not to leave them bottleneck as well. Furthermore, 
%and perform optimizations in HLS with respect to scheduling, resource allocation, clock frequency and memory usage. We analyze the dependencies within the ModMult algorithm and generating a tight scheduling that gives a near-optimal clock cycles. Our framework is targeted at datacenter acceleration cards, whose peak performance is upper-bounded by the number of DSPs\cite{parker2014understanding}. Therefore, we optimize the resource usage without sacrificing the latency, and also adopt the Karatsuba algorithm to construct multipliers that make a conserve use of DSP blocks. The resulted Paillier cores are compact and scalable, with a DSP-efficiency outperforming the existing works.

The hardware module are built as OpenCL kernels and incorporate into FATE as an encryption library.
Each kernel performs en/decryption for a batch of data to relieve the kernel invocation overhead, and kernels are queued in the OpenCL command queue to help overlap data transfer with computation and hide latency. The proposed encryption framework is general and does not require any change of the model, while preserving the security and accuracy.

We perform extensive evaluation on the proposed framework, demonstrating that it reduces the iteration time for training linear models by up to 26\%, and the encryption time in each iteration by 71\%. Our hardware framework delivers an acceleration ratio of 10.6 for encryption and 2.8 for decryption compared with software solutions. Our circuit for ModMult operation achieves a better DSP-efficiency than existing FPGA solutions, with a comparable execution latency but a lower usage of DSP blocks.

We summarize our contributions as follows.

\begin{itemize}
	\item Introducing a hardware-based encryption framework for federated learning, achieving high efficiency without sacrifice of security and utility, supporting accelerated computation in cloud datacenters.
	\item Presenting architectures for Paillier homomorphic cryptosystem taking a scalable approach making efficient utilization of the FPGA resources, especially DSP blocks.
	\item Incorporating the encryption framework into cutting-edge federated learning framework, and showing an improvement on training throughput for federated learning models.
\end{itemize}

The rest of the article will be organized as follows. 
In Section \ref{sec:rel} we will provide the background about federated learning and existing privacy preserving machine learning systems, and introduce the Paillier cryptosystem we work on. Section \ref{sec:des} will present the design and implementation of the framework in detail. Section \ref{sec:eval} shows the methodology and results of evaluation. Finally in Section \ref{sec:con} we conclude the article.

\section{Background}\label{sec:rel}

\subsection{Federated learning with HE}
Federated learning is a privacy-preserving, decentralized distributed machine learning paradigm. One effective method of preserving privacy and securing computation is homomorphic encryption (HE), i.e.\ encryption schemes that allows encrypted values to be involved in computation. For the applications of HE in federated learning, we refer the readers to \cite{hardy2017private}, \cite{gilad2016cryptonets}, \cite{aono2017privacy}, \cite{liu2019privacy}, \cite{liu2018secure}, \cite{chai2019secure}, which broadly cover machine learning models including linear model, neural network and deep learning, boosting tree, transfer learning and matrix factorization. Typically, HE is employed to encrypt the intermediate data during computation, which will then be transferred and aggregated by homomorphic operation. For nonlinear operations composing the model, such as activation function in a neural network, these works usually rely on approximation to make the model agree with HE computation.

\subsection{Privacy-preserving ML systems}

There has also been machine learning systems that take privacy preservation into account, such as SecureML\cite{mohassel2017secureml} that proposes a system for two non-colluding party collectively training a model, and Sage\cite{Sage} that presents a differentially private machine learning platform. Among them, FATE\cite{FATE} introduces a federated learning framework that provides the abstraction and utilities for implementing algorithm and models, along with an architecture to enable distributed, multiparty machine learning. It mainly utilizes Paillier homomorphic encryption to guarantee data security. However, it purely relies on a software solution of encryption that greatly harms the execution efficiency of federated training. Our goal in the work is to find a hardware solution as a rescue to this issue.

%\subsection{Hardware acceleration of HE}

%(Supplement later)

%\section{Priliminaries}\label{sec:pri}

\subsection{Paillier Homomorphic Encryption}

\begin{figure*}[th]
	\centering
	\includegraphics[width=0.75\linewidth]{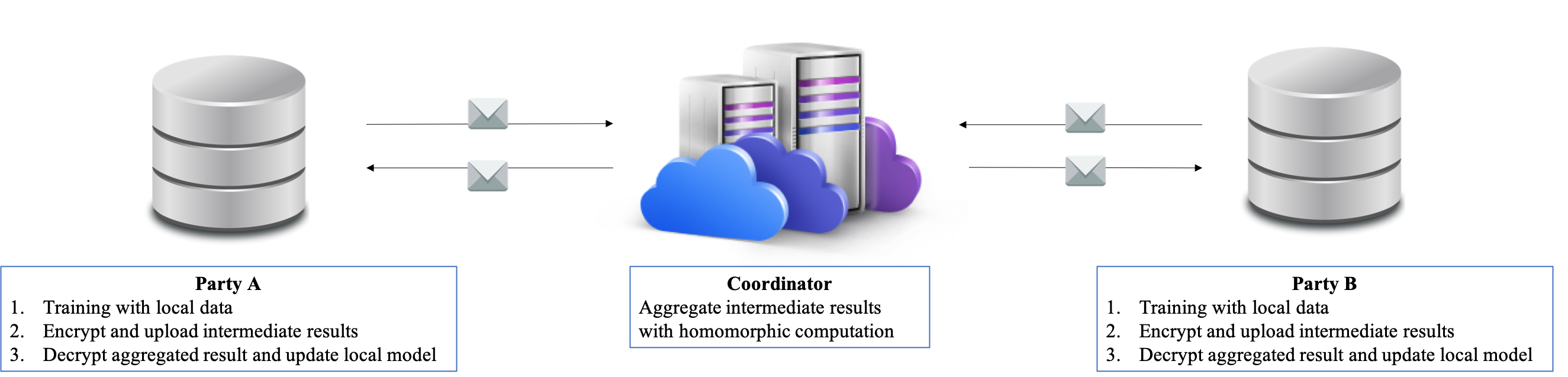}
	\caption{General workflow of homomorphic encryption-based federated learning}
	\label{fig:homo}
\end{figure*}
Paillier HE is an additive homomorphic encryption scheme allowing to perform addition and multiplication with scalar on encrypted values without decrypting them. 
In federated learning, usually multiple parties are involved, each one having a private dataset and wanting to maintain a local model learned from the aggregated dataset, and there may be a coordinator to manage the computation and data exchange among parties (Figure \ref{fig:homo}). The role of Paillier homomorphic encryption is to encrypt the intermediate data to transfer, so that in each training iteration the coordinator receives the encrypted local updates from parties, aggregates them with the homomorphic property, and sends back the result to each party for decryption and updating local model. In this way, each party obtain a model extracting information from the aggregate dataset, without leaking its private information.

The Paillier HE scheme associates each party with a public key $(n, g)$ and a private key $(\lambda, \mu)$, where $n, g, \lambda, \mu$ are large integers, typically 1024-bit in FATE. Messages and ciphertexts are also represented as long integers. A message $m$ can be encrypted into ciphertext $c$ by $c=g^mr^n\mod n^2$ with random number $r$, and decryption is performed by $m=((c^\lambda \mod n^2)-1)/n*\mu\mod n^2$.

We can see from the formulation that the majority of the computation of the Paillier en/decryption is related to modular exponentiation (ModExp), which can be further decomposed to a series of ModMult operations. Hence, the execution of ModMult has a decisive effect on the overall performance. We choose the Montgomery ModMult algorithm\cite{montgomery1985modular} to perform this operation because it is FPGA-friendly, in that it disposes of the costly integer division. The Montgomery algorithm, shown in Algorithm \ref{alg:mul_word}, computes $XY\cdot 2^{-l}\mod M$ for $l$-bit integers $X$, $Y$ and $M$. It divides integers into $k$-bit words. The body of the algorithm is a two-level loop, where each outer iteration (line 2-8) aims to compute an intermediate result $S_i=X\cdot Y^i \cdot 2^k \mod M$ for the $i$th word of $Y$, and it further decomposes the computation by each word of $X$ and forms the inner loop (line 4-6).

\begin{algorithm}[ht]
	\KwIn{$X=\sum_{j=0}^{l/k-1}X^j\cdot 2^{jk}$, $Y=\sum_{j=0}^{l/k-1}Y^j\cdot 2^{jk}$,  $M=\sum_{j=0}^{l/k-1}M^j\cdot 2^{jk}$, $r=2^k$}
	\KwOut{$S = X \cdot Y / 2^{l} \mbox{ mod } M$}
	$S_0\la 0$\;
	\For{$i=0\ldots l/k-1$}{
		$q \la ((S_i + X*Y^i)\cdot (-M^{-1})) \mbox{ mod }r$\;
		
		\For{$j=0\ldots l/k$} {
			$\bar{S}_{i+1}^j \la S_{i}^j + X^j*Y^{i} + q * M^j$\;
		}
		$S_{i+1} \la \bar{S}_{i+1} / 2^k$
	}
	\If{$S_{l/k} > M$}{$S_{l/k} \la S_{l/k} - M$\;}
	\KwRet{$S_{l/k}$}
	\caption{Montgomery Algorithm for Modular Multiplication with Radix $2^k$}
	\label{alg:mul_word}
\end{algorithm}

\section{Design and Implementation}\label{sec:des}

\subsection{System Overview}

\begin{figure}[th]
	\centering
	\includegraphics[width=0.8\linewidth]{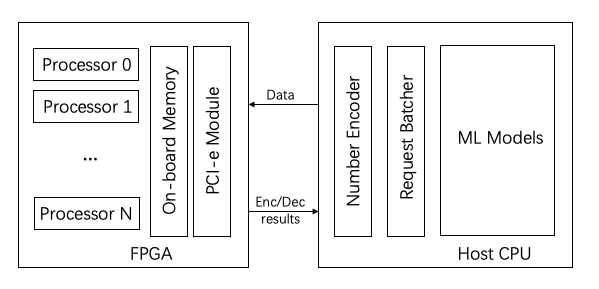}
	\caption{Overview of Our Encryption Framework}
	\label{fig:overview}
\end{figure}

The overall architecture of our encryption framework is shown in Figure \ref{fig:overview}. The framework is envisioned to be hosted on cloud servers belonging to geo-distributed parties of federated learning. It includes components residing on both the host CPU and the FPGA, where a PCI-e bus provides communication between them.
The host CPU is responsible for the normal training workload of a machine learning model, while it batches the requests of encryption to sends to the FPGA, and encodes the floating point number used by machine learning to integers agree with HE schemes.
Apart from these necessities, our main contribution is designing high performance processors for Paillier computation on FPGA and encapsulating the hardware module as OpenCL kernel for invocation, which we will detail in Section \ref{ssec:micro} and \ref{ssec:impl} respectively.

\subsection{Micro-architecture for Montgomery ModMult}\label{ssec:micro}
A Paillier processor encapsulates units for operations involved, i.e.\ modular multiplication, random number generator and integer divisor, along with its local storage. We replicate Paillier processors in HLS to deploy multiple copies, and the top level function is responsible for dispatching input data and collecting results. Since the Paillier processors are independent and work in parallel, the overall throughput of an FPGA chip can be determined by 
$$
\mathrm{Throughput} = \frac{\mbox{Total amount of resource}}{\mbox{Latency} \times \mbox{Resource consumption per core}},
$$
where resource broadly refers to multipliers, adders, memory, etc., and latency can be further decomposed to clock cycle of execution $\times$ clock frequency. Therefore, our design guideline is to optimize the Montgomery ModMult operation lying at the heart of Paillier cryptosystem, with respect to clock cycle, resource allocation, clock frequency, in addition to memory usage. We elaborate on the optimization on these dimensions as below.

\subsubsection{Clock Cycle}
Generally, the clock cycle required by an algorithm is intrinsically lower bounded by the number of operations and the critical path in the dependency graph. As shown in Algorithm \ref{alg:mul_word}, the body of the Montgomery algorithm is a two-level loop, consisting of $2(l/k)(l/k+1)$ multiplications. Thus, the ideal clock cycle will be $2(l/k)(l/k+1)$ divided by the number of multipliers, even if we ignore the rest of the operations. On the other hand, as the execution of each inner iteration depends on the iteration before, it is hard to force a parallel execution of the inner iterations. Our goal is to deploy two multipliers for an inner iteration, and obtain a clock cycle number as close to $(l/k)(l/k+1)$ as possible.

Another dependency issue that deserves attention is the computation of $q$ in each outer iteration. In the $i$th iteration, $q$ depends on the value of $S_{i-1}$, while it is necessary in the computation of inner loops. However, if $q$ is computed before the start of inner loop, the latency will be magnified by the number of outer iterations. 

To enforce a tight scheduling, we make the following optimizations in HLS:

\begin{itemize}
	\item Unrolling the inner loop. This can be done through an UNROLL directive in HLS, or manually repeat part of the loop. Unrolling the loop does not lead to parallel execution of iterations. However, this is the only way to disassemble all the operations composing the loop, achieving the flexibility of scheduling to overlap operations as much as possible. Also, without unrolling we are not able to insert the computation of $q$ into the middle of an inner loop.
	\item Interleaving the $q$ computation with the inner loop. As discussed before, the $q$ value used in each inner loop must be computed before. Since the $q$ value for computing $S_i$ only relies on first few words of $S_{i-1}$, it is possible to start generating $q$ in the last inner loop when those words are ready. In this way we can obtain $q$ in advance and hide its latency.
	\item Pipelining the outer loop. We achieve this by inserting a PIPELINE directive in HLS, with the initiation interval set to the number of iterations contained in an inner loop. The final step of schedule enforcement is to pipeline all the iterations. We aim to pipeline both the outer loop and the inner loop by unrolling the inner loop, and pipelining the outer loop, so that the inner loop is naturally pipelined by scheduling the disassembled operations, and the outer loop try to start an interaction each time when a whole inner loop is initiated.
	
\end{itemize}

\begin{figure*}[ht]
	\begin{tabular}{cc}
		\begin{minipage}[t]{0.70\linewidth}
			\includegraphics[width = 1\linewidth]{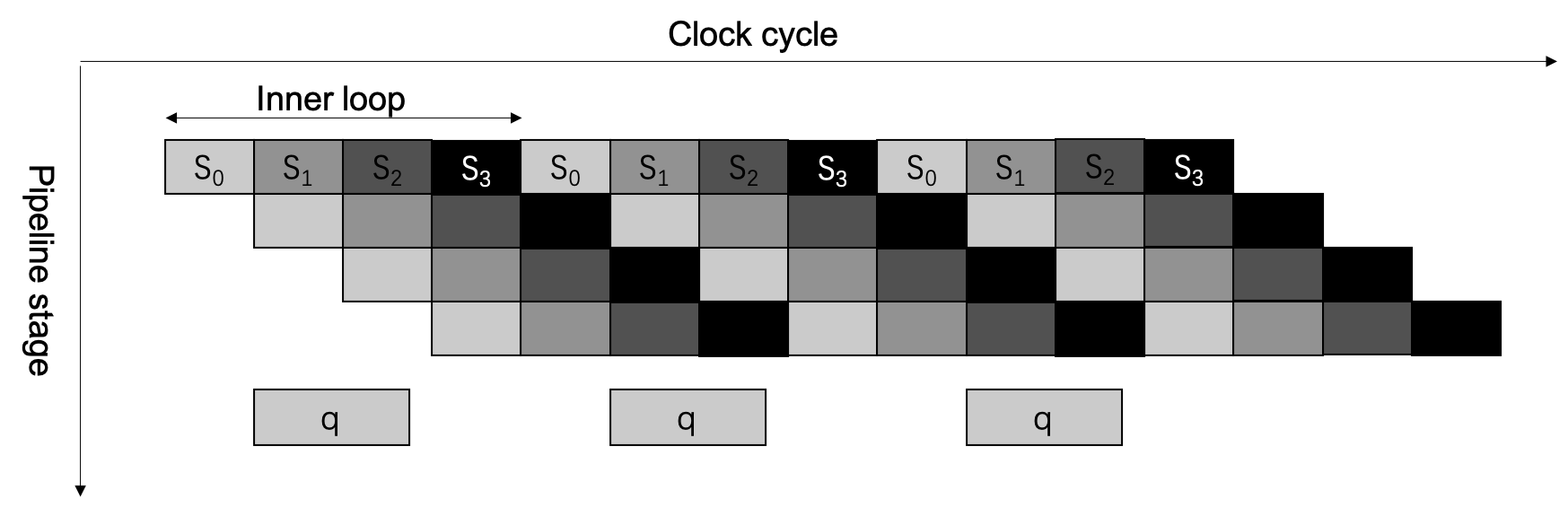}
			\caption{Pipeline Execution of the Montgomery ModMult Operation}
			\label{fig:sched}
		\end{minipage}
		\begin{minipage}[t]{0.28\linewidth}
			\includegraphics[width = 1\linewidth]{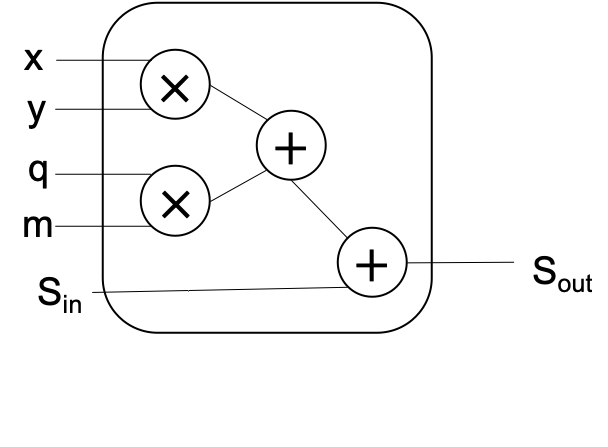}
			\caption{Processing Element Implementing the Inner Loop of Algorithm \ref{alg:mul_word}}
			\label{fig:pe}
		\end{minipage}
	\end{tabular}
	\label{fig:design}
\end{figure*}

%\begin{figure*}[th]
%	\centering
%	\includegraphics[width=0.9\linewidth]{sched.png}
%	\caption{Pipeline Execution of the Montgomery ModMult Operation}
%	\label{fig:sched}
%\end{figure*}

The resulted scheduling is shown in Figure \ref{fig:sched}. We illustrate with an example with operands 4 words in length (i.e.\ $l/k=4$), and the computation of each inner iteration takes 4 clock cycles to complete. Initially, the schedule computes $q$ for the first inner loop (not shown in the figure), and then initiates the inner iterations sequentially. In the meantime, as soon as $S^0$ is ready, it can be used to compute $q$ for the next inner loop. Hence, when the last inner iteration ends, the first iteration of the next inner loop can start immediately with the precomputed $q$. Therefore, we enable a tight schedule that initiates an inner iteration each clock cycle. The resulted execution clock cycle is $(l/k)(l/k + 1)$, plus the number of pipeline stages, and a few cycles for data read-in and write-out.

\subsubsection{Resource Allocation}
In this work, we utilize the embedded DSP blocks on the FPGA chip to construct pipelined multipliers. For the remaining logic, including adder, multiplexer, integer comparison, finite state machine, etc., we leave them to lookup-table (LUT). As DSP on FPGA is scarce and expensive, we use them to carry out the heavy multiplication only. Further, we will show purely relying on LUT to implement ModMult operation is not economic (Section \ref{sec:eval}). Therefore, we will focus on the usage of LUT and DSP, and reduce the area and DSP usage without sacrificing the performance.

We encapsulate the operations comprising an inner iteration into a processing element (PE), as shown in Figure \ref{fig:pe}. Each PE contains two multipliers to perform the two independent multiplications $x*y$ and $q*m$. Then it accepts $S_{i-1}^j$ of the last outer iteration, and a carry word (not shown in the figure), adds them with the multiplication results, and then outputs $S_{i}^j$ and a carry word. Then we limit the number of PE to 1, with an ALLOCATION directive in HLS. This is to avoid the resource bloating owing to loop unrolling, so that only resource for computing one inner iteration is actually allocated, and to reduce the overall area of the micro-architecture.

We also employ the Karatsuba algorithm to construct DSP-conservative multipliers. As shown in Algorithm \ref{alg:karatsuba}, Karatsuba algorithm performs an integer multiplication by recursively breaking it into three of half size. Its efficiency is attributed to one multiplication less than the schoolbook algorithm, and we take advantage of it to allocate DSPs according to the actual number of operations. For instance, a DSP48E1 block is able to carry out $18\times 25$-bit multiplication, and a $32\times 32$ one can be divided into $16\times 16$ ones, and takes up 3 DSP blocks.

\begin{algorithm}[ht]
	\KwIn{Operands $X$ and $Y$, the length of operand $k$}
	\KwOut{$S=X*Y$}
	Let $X=\overline{X_hX_l}$, $Y=\overline{Y_hY_l}$, where $X_h, X_l, Y_h, Y_L$ are $k/2$-bit integers\;
	$HH\la Karatsuba(X_h, Y_h)$\;
	$LL\la Karatsuba(X_l, Y_l)$\;
	$HL\la Katatsuba(X_h + X_l, Y_h + Y_l)$\;
	$S\la HH * 2^k + (HL - HH - LL) * 2 ^{k/2} + LL$\;
	\KwRet{$S$}
	\caption{Karatsuba algorithm}
	\label{alg:karatsuba}
\end{algorithm}

%\begin{figure}[th]
%	\centering
%	\includegraphics[width=0.8\linewidth]{pe.png}
%	\caption{Processing Element Implementing the Inner Loop of Algorithm \ref{alg:mul_word}}
%	\label{fig:pe}
%\end{figure}

\subsubsection{Clock Frequency}
The DSP units on the Xilinx FPGA run at a maximum frequency of 400-500MHz. To approach the frequency limit, we need to pay attention to the following measures:

\begin{itemize}
	\item Declare the multipliers as pipelined multipliers. A pipelined multiplier takes multiple cycles to accomplish a multiplication, distributing its workload and relieving the burden of each cycle. It does no harm to the multiplication throughput since we have resolved the dependency between its input and output.
	\item Restrict bitwidth of operands. The clock frequency is constraint by the critical path of the circuit, i.e.\ the longest path of gates a signal needs to pass through during one cycle. Arithmetic on integers such as addition or comparison usually results in a long carry chain, and thus we need to avoid computation on very long integers directly. In this work, we use 32-bit as the operand size, and the maximum bitwidth involved is 64 bits.
	\item Simplify the control logic. For the finite state machine in charge of controlling the compute units, we use one-hot encoding scheme to represent the states for a fast lookup and match. The number of states is related to the number of iterations of each loop and thus one-hot encoding will be acceptable.
\end{itemize}

\subsubsection{Memory Usage}
Our design allocate each Paillier processors its own block RAM (BRAM) as local buffer, to hold the input/output data and the intermediate large integers involved in the computation. We do not share storage among processors to prevent data access contention.

Large integers are normally stored as arrays of words in the BRAM. However, we notice that the input data for encryption, which are encoded from floating point numbers used in machine learning, have few effective digits compared with the length of large integers. Therefore, we are able to store the input data as a sparse vector, i.e.\ only recording the non-zero elements and their indices, reducing the memory footprint.

\subsection{Implementation}\label{ssec:impl}
We develop our encryption framework with the AWS F1 instance and Xilinx SDAccel development suite. The basic logic of the encryption and decryption function is implemented with Xilinx high level synthesis (HLS), allowing to transform an algorithm described in C/C++ into tailor-made implementation on FPGA. Directives like loop pipelining and instance allocation are inserted into the HLS code to fine tune the performance of the resulted architecture.

On the host side, we use the OpenCL API to access the acceleration hardware. The OpenCL API provides an abstraction of the computing device like CPU, GPU and FPGA. An invocation to the device function is named a kernel. OpenCL is used to manage the data transfer between host and device, queue and invoke kernels, and monitor the execution events. We adopt the PyOpenCL APIs to implement a module that makes use of the FPGA device for cryptographic processes and incorporate it into the FATE framework.

The requests from the host side are divided into fixed size batches, and each batch invokes a kernel on device. Multiple kernels will be queued in the OpenCL command queue. This helps overlap data transfer with computation and hide latency.
We also preallocate buffers on the device, arranging them as a ring buffer, in order to reuse buffers among kernels and avoid frequent memory allocation.

\begin{figure}[th]
	\centering
	\includegraphics[width=0.8\linewidth]{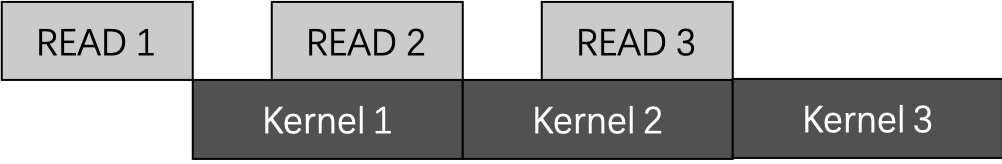}
	\caption{Queueing kernels for execution}
	\label{fig:cmdQ}
\end{figure}

\section{Evaluation}\label{sec:eval}
We conduct experiments aiming to perform an extensive evaluation on the proposed encryption framework. We first perform a microscopic examination, comparing the implementation of Paillier algorithm and ModMult operation with software solutions and existing FPGA designs. Then we study its improvement on the overall performance of training process of federated learning. The training tasks are carried out on the open-sourced version of the FATE machine learning framework. We choose two linear models, and adopt Kaggle datasets on breast cancer\footnote{https://www.kaggle.com/uciml/breast-cancer-wisconsin-data} and motor temperature\footnote{https://www.kaggle.com/wkirgsn/electric-motor-temperature} and partition the datasets vertically.

We attempt to answer the following questions empirically with the evaluation experiments:

\begin{itemize}
	\item How do the Paillier processors perform, especially for the ModMult operation, in terms of throughput and resource-efficiency?
	\item How does the hardware framework compare with software solutions of Paillier cryptosystem in terms of en/decryption throughput?
	\item How much does the framework affect the training throughput of federated learning with respect to different models or algorithms?
\end{itemize}

\begin{table*}[h]
	\centering
	\begin{tabular}{cccccc}
		\toprule
		Implementation & Area(slice) & DSP & Clock frequency(MHz) & Execution time(us) & Throughput per DSP(op/s) \\ 
		\midrule 
		This work & 483 & 9 & 500 & 8.81 & {\bf 12626} \\ 
		\hline 
		\cite{SanA14} & 567 & 13 & 490 & 8.64 & 8903 \\ 
		\hline 
		\cite{SongKNI10}	 & 180 & 1 & 447 & 135.4 & 7385 \\ 
		\hline 
		\cite{HuangGE11}	 & 9268 & NA & 129 & 18.70 & NA \\ 
		\hline 
	\end{tabular}
	\caption{Comparison of ModMult operaion}\label{tab:mm}
\end{table*}

Given the broad adoption of the ModMult operation, many implementation has been proposed by researchers, and we compare ours with them in Table \ref{tab:mm}. Since we are targeting datacenter acceleration chips and applications, the DSP efficiency is a key factor evaluating an implementation. Comparing with the state of the art solution \cite{SanA14}, our ModMult module delivers a close latency but uses fewer DSPs due to our precise limit on resource usage. The authors of \cite{SongKNI10} proposes an implementation using only one DSP and one block RAM. However, without employing the Karatsuba algorithm, their version turns out to be less efficient than ours. \cite{HuangGE11} gives an implementation using circuit elements entirely without DSP, and it shows that an such a ModMult module consumes much area and limits the clock frequency, and hence not recommendable. Moreover, most of existing solutions are based on register-transfer level (RTL) that describes the circuit directly, but lacks the flexibility of parametrizing and reusing the ModMult module as our HLS version does. 

\begin{figure}[ht]
	\begin{tabular}{cc}
		\begin{minipage}[t]{0.48\linewidth}
			\includegraphics[width = 1\linewidth]{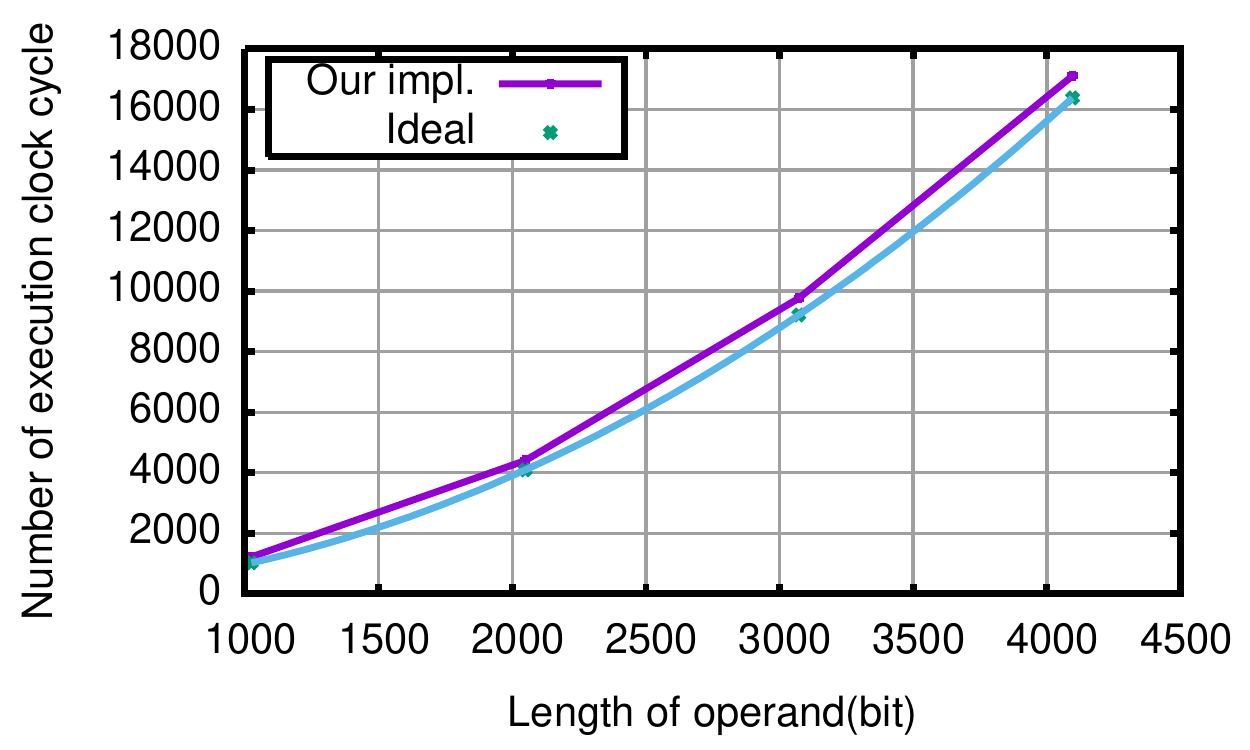}
			\caption{Number of execution clock cycles of ModMult operation}
			\label{fig:cc}
			%			\caption{This figure shows the performance of retrieval from different methods.}
		\end{minipage}
		\begin{minipage}[t]{0.48\linewidth}
			\includegraphics[width = 1\linewidth]{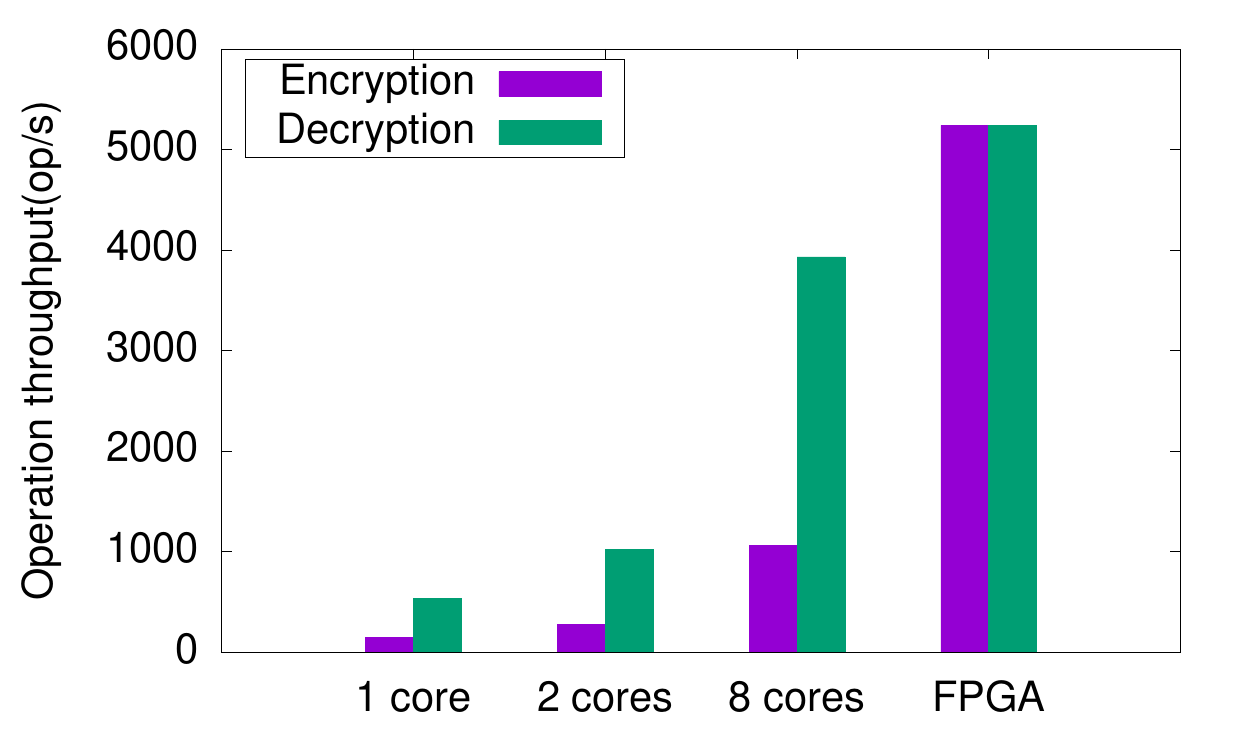}
			\caption{Throughput of FPGA and multicore processor}
			\label{fig:mc}
			%			\caption{This figure shows the performance of different methods in other evaluations.}
		\end{minipage}
	\end{tabular}
	%    \caption{Encryption and decryption throughput}
	\label{fig:tmp}
\end{figure}

To evaluate the effectiveness of the scheduling of ModMult operation, we compare the number of execution clock cycles with the theoretically ideal clock cycle, given as $T=(l/k)(l/k+1)$ (Section \ref{sec:des}). As shown in Figure \ref{fig:cc}, for different sizes of operands, our implementation keeps no more than 10\% higher than the ideal. The gap is mainly due to pipeline stages, time for initialization and data transfer.

%\begin{figure}[th]
%	\centering
%	\includegraphics[width=0.8\linewidth]{cc.pdf}
%	\caption{Number of execution clock cycles of ModMult operation}
%	\label{fig:cc}
%\end{figure}

%\begin{figure*}[!h]
%	\centering
%	\subfigure[Encryption]{
%		\includegraphics[width=0.22\textwidth]{phe1.pdf}
%		\label{fig:phe1}
%	}
%	\subfigure[Decryption]{
%		\includegraphics[width=0.22\textwidth]{phe2.pdf}
%		\label{fig:phe2}
%	}
%	\caption{Encryption and decryption throughput}
%	\label{fig:phe} %% label for entire figure
%\end{figure*}

\begin{figure}[ht]
	\begin{tabular}{cc}
		\begin{minipage}[t]{0.48\linewidth}
			\includegraphics[width = 1\linewidth]{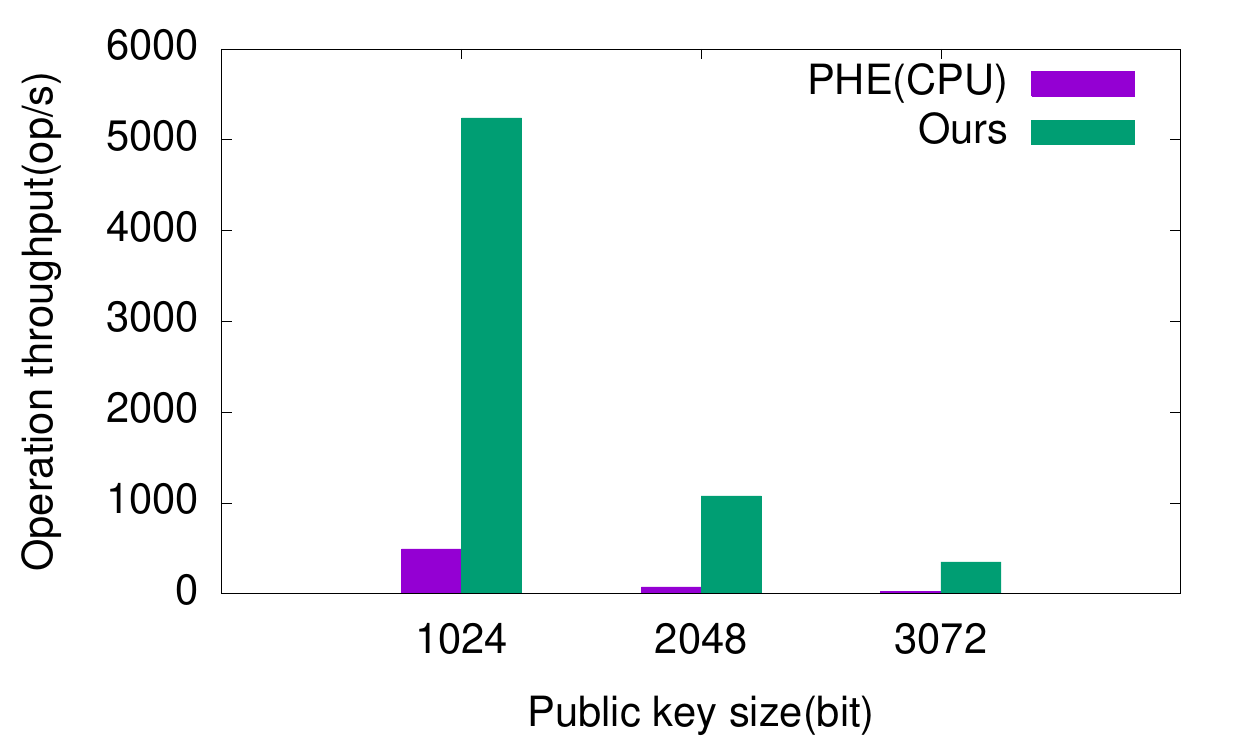}
			\caption{Encryption Throughput Compared with Software}
			\label{fig:phe1}
%			\caption{This figure shows the performance of retrieval from different methods.}
		\end{minipage}
		\begin{minipage}[t]{0.48\linewidth}
			\includegraphics[width = 1\linewidth]{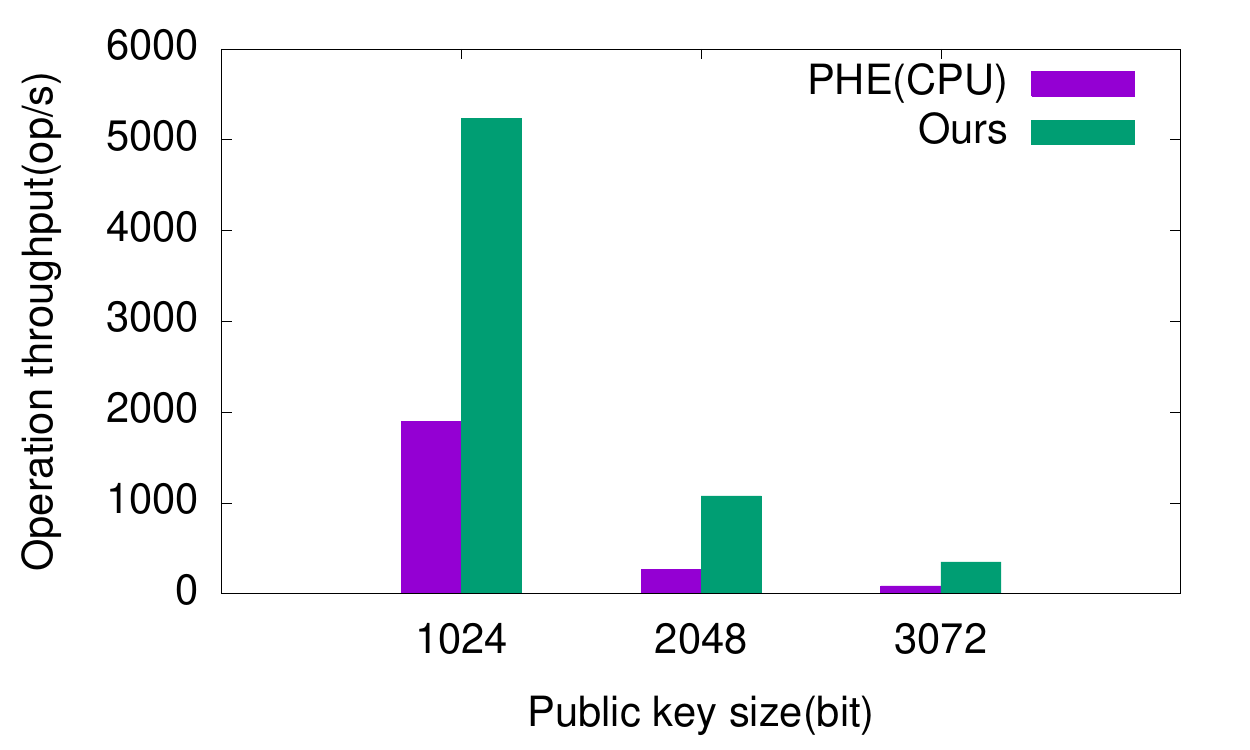}
			\caption{Decryption Throughput Compared with Software}
			\label{fig:phe2}
%			\caption{This figure shows the performance of different methods in other evaluations.}
		\end{minipage}
	\end{tabular}
%    \caption{Encryption and decryption throughput}
    \label{fig:phe}
\end{figure}

To investigate the performance of FPGA and software solution, we compare the framework with PHE, a popular Paillier library, as shown in Figure \ref{fig:phe1} and \ref{fig:phe2}. We can see that for a 1024-bit public key, our framework delivers an acceleration ratio of $10.62\times$ and $2.76\times$ for encryption and decryption, respectively. We also compare FPGA with a multicore processor using {\tt libpaillier} library, as shown in Figure \ref{fig:mc}. It shows that an FPGA effectively outperforms a multicore CPU and is advisable to be used in accelerating computational intensive applications.

\begin{figure}[ht]
	\begin{tabular}{cc}
		\begin{minipage}[t]{0.48\linewidth}
			\includegraphics[width = 1\linewidth]{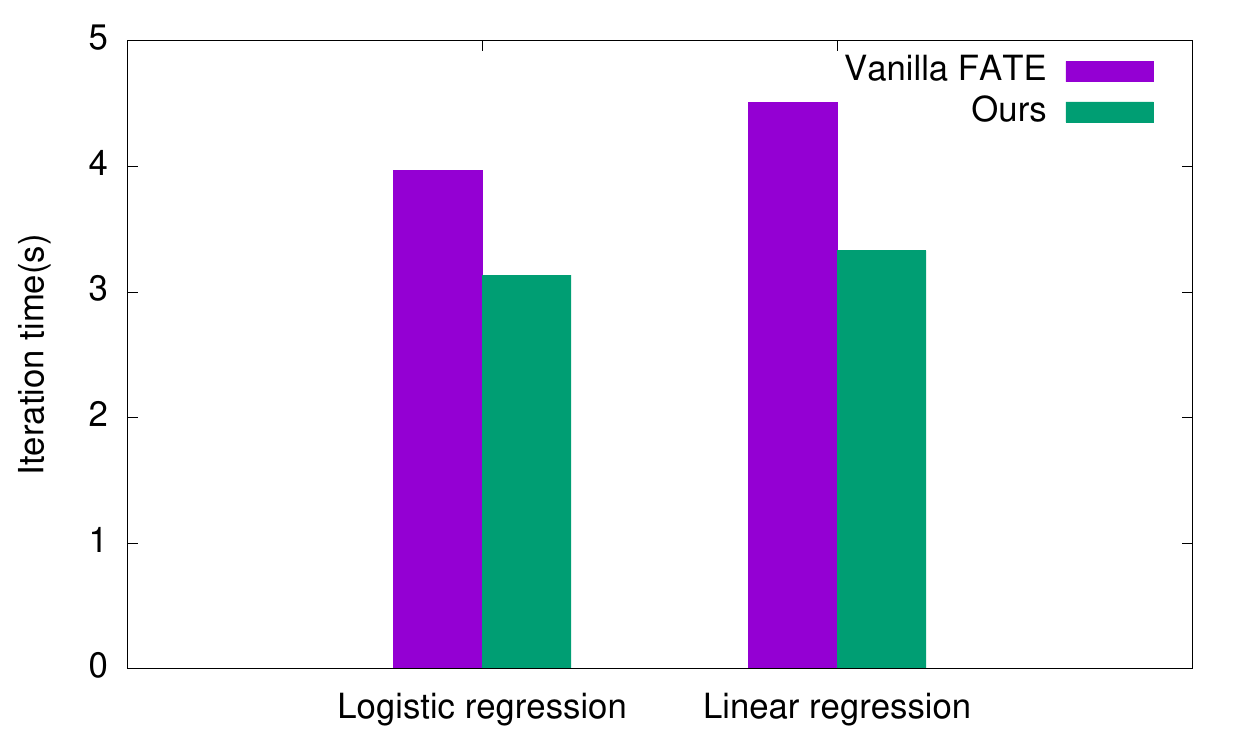}
			\caption{Improvement on Iteration Time}
			\label{fig:fate1}
			%			\caption{This figure shows the performance of retrieval from different methods.}
		\end{minipage}
		\begin{minipage}[t]{0.48\linewidth}
			\includegraphics[width = 1\linewidth]{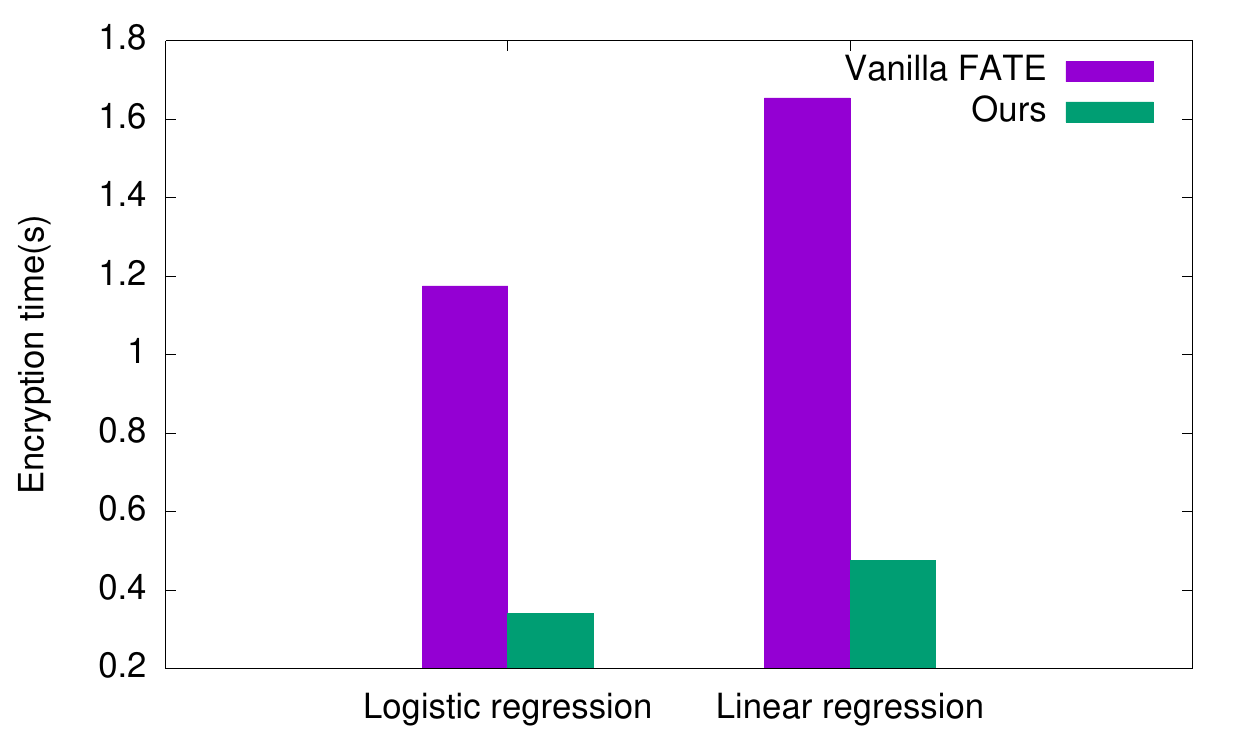}
			\caption{Improvement on Encryption Time Per Iteration}
			\label{fig:fate2}
			%			\caption{This figure shows the performance of different methods in other evaluations.}
		\end{minipage}
	\end{tabular}
	%    \caption{Encryption and decryption throughput}
	\label{fig:fate}
\end{figure}

Additionally, we test the modified FATE with linear models and the breast and motor datasets. We train a logistic regression and a linear regression model on the two datasets respectively for 10 iterations, and record the timing.
Figure \ref{fig:fate1} and Figure \ref{fig:fate2} show the training iteration time and the encryption time in each iteration respectively. It demonstrates that for linear models, our framework reduce the training iteration time by up to 26\%, and the encryption time during one iteration by 71.2\%.

%\begin{table*}[h]
%	\centering
%	\begin{tabular}{cccccc}
%		\toprule
%	Implementation & Area(slice) & DSP & Clock frequency(MHz) & Execution time(us) & Throughput per DSP(op/s) \\ 
%		\midrule 
%	This work & 483 & 9 & 500 & 8.81 & {\bf 12626} \\ 
%		\hline 
%	\cite{SanA14} & 567 & 13 & 490 & 8.64 & 8903 \\ 
%		\hline 
%	\cite{SongKNI10}	 & 180 & 1 & 447 & 135.4 & 7385 \\ 
%		\hline 
%	\cite{HuangGE11}	 & 9268 & NA & 129 & 18.70 & NA \\ 
%		\hline 
%	\end{tabular}
%\caption{Comparison of ModMult operaion}\label{tab:mm}
%\end{table*}

%\begin{figure}[th]
%	\centering
%	\includegraphics[width=0.8\linewidth]{mc.pdf}
%	\caption{Throughput of FPGA and multicore processor}
%	\label{fig:mc}
%\end{figure}

%\begin{figure*}[ht]
%	\centering
%	\subfigure[Iteration time]{
%		\includegraphics[width=0.4\textwidth]{fate1.pdf}
%		\label{fig:fate1}
%	}
%	\subfigure[Encryption time]{
%		\includegraphics[width=0.4\textwidth]{fate2.pdf}
%		\label{fig:fate2}
%	}
%	\caption{Acceleration of federated learning model}
%	\label{fig:fate} %% label for entire figure
%\end{figure*}

\section{Conclusion}\label{sec:con}
In this paper, we have demonstrated the significance of accelerating homomorphic encryption and modular operations. We explored a compact architecture for Paillier cryptosystem with an HLS-based approach, investigating how to optimize the performance, and incorporated the FPGA framework into a federated learning system. We conducted extensive experiments to present the effectiveness and efficiency of our encryption framework.

\newpage
%\nocite{*}
\bibliographystyle{named}
\bibliography{ijcai20}

\end{document}